\documentclass[preprint,12pt]{aastex}
\usepackage{verbatim}
\usepackage{epsfig, amsmath, amsfonts}
\usepackage{natbib}

\begin{document}

\title{Jacquetium: a new, naturally-occurring chemical element}
\author{Emmanuel Jacquet\\
\textit{Jacquet Institute of Eclectic Science in Buc}\\
}

\begin{abstract}
I report the discovery of jacquetium ($_0$Jq), the first naturally occurring element found since more than 80 years. It is volumically the most important element of the local Universe.
\end{abstract}

\textit{Received April 1, 2019; accepted in revised form April 1, 2025}

\section{Introduction}
Since, 150 years ago, \citet{Mendeleev1869} devised his periodic table, the historical search for new chemical elements narrowed its scope to definite atomic numbers $Z$. By the 1940s, all elements between hydrogen ($_1$H) and uranium ($_{92}$U) had been found, and, save for technetium, identified in nature including meteorites (by the 1950s for Xe and Kr; \citealt{Mason1962}). While transuranian elements have been discovered since, few nuclides are stable over appreciable timescales, and only neptunium, plutonium, and the recently evidenced curium \citep{Brenneckaetal2010} demonstrably exist in nature, if in too minute amounts to be industrially useful \citep[see e.g.][]{Brown1985}. Philosophically, the question arises whether the recognition of chemical elements without proven natural occurrence, at variance with standard (International Mineralogical Association) practice for mineral species, is meaningful in any way. At any rate, it is understandable that, notwithstanding theoretical hopes for an island of relatively stable superheavy isotopes \citep[e.g.][]{Cooper2013, Hofstadter2013}, the focus in physics has largely shifted to new elementary particles (e.g. \citealt{Andari2012}; but see \citealt{Fekete2019}).

  In this context, the significance of the discovery of a new \textit{natural} element is hard to overstate. In this paper, I report the discovery of Jacquetium, the first naturally-occurring chemical element found since more than 80 years, and the first such one in the New Millenium. It is also the last \textit{cisuranian} ($Z\leq 92$) element ever, hereby completing a long search by generations of chemists. I now set to describe its proposed name and properties.

\section{Element name}
 The name ``Jacquetium'' (symbol Jq) proposed for the new element described herein honors Dr Emmanuel Jacquet, a renowned and therefore French cosmochemist. Beside a notable tendency to self-citation 
 and three ping-pong championship titles at the Museum national d'Histoire naturelle mineralogy laboratory (2012, 2013, 2019), his chief claim to fame lies in his successful prediction of the evenness of the Avogadro number (\citet{JacquetNA}; see 26th conference of International Board  of Weights and Measures), and his pioneering geochemical study on the isotopic fractionation of sodium \citep{JacquetF}.

\section{Properties of Jacquetium}
Jacquetium is the element characterized by the atomic number $Z=0$ ($_0$Jq). Its existence is easily proven by several naturally-occurring isotopic forms:

Jacquetium-1 is the free ``neutron'' \citep{Chadwick1932}. It occurs in nature as a transient agent of stellar \citep[e.g.][]{Lugaroetal2018} and interplanetary \citep[e.g.][]{HerzogCaffee2014} nucleosynthesis
. It is nonetheless unstable and undergoes $\beta$-decay with a half-life of 10 minutes according to the reaction:
\begin{equation}
_0^1\mathrm{Jq}\rightarrow _1^1\mathrm{H}+\beta^{-} +\overline{\nu}
\end{equation}

Unlike jacquetium-1, jacquetium-0 appears very stable and is volumically the most important atom of the local universe. It may be identified with the newtonium which \citet{Mendeleev1904} himself had postulated to constitute the ether. Yet some decay over billion-years timescales might incur some variation in the cosmological constant.

Underneath their crusts, neutron stars may be largely viewed as single atoms of jacquetium, specifically jacquetium-10$^{57}$, and as such prove the existence of an island of stable superheavy isotopes.

\begin{figure}
\resizebox{\hsize}{!}{
\includegraphics{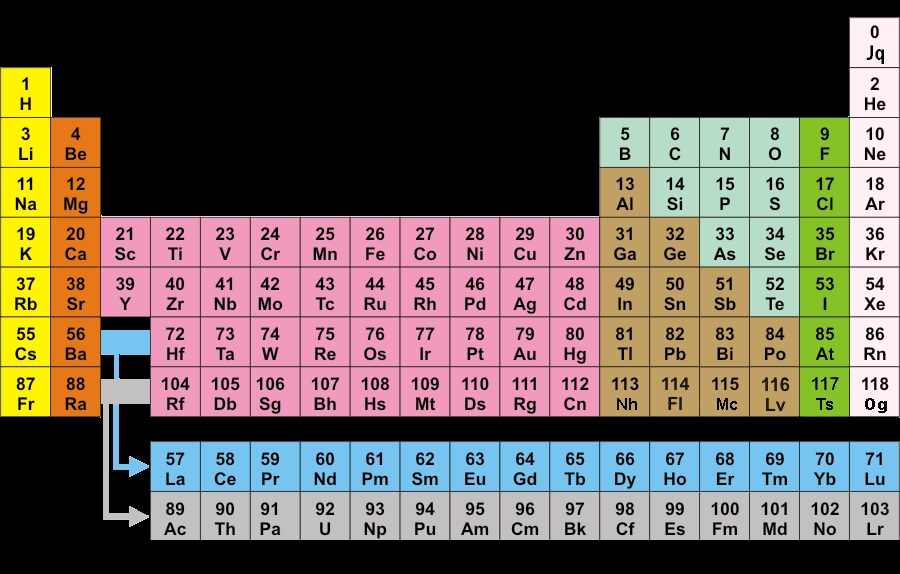}
}
\caption{Revised periodic table with the new element jacquetium. Copyright Tommy Vettor.}
\label{periodic}
\end{figure}

Despite occasional formation of an anion $^0$Jq$^-$ also known as the free electron, lack of molecular combination of jacquetium with other chemical elements warrants its inclusion among noble gases, above helium in the periodic table (Fig. \ref{periodic}). Although IUPAC recommends new noble gas names to end with "-on" \citep{Corish2016}, the "-ium" ending chosen aligns well with the next noble gas "helium". Still, "jacqueton" is also undeniably euphonious and should be bestowed on the next discovered elementary particle (Jacquet, in progress).

\section{Conclusion}
I have proven beyond doubt the existence of a new element, herein christened jacquetium, with $Z=0$, of which the neutron is but one isotope. I therefore urge all science classrooms in the world to update their periodic tables and add ``Jq'' at the top of the noble gas column (see Fig. \ref{periodic}). The discovery of a new natural element 150 years after the landmark \citet{Mendeleev1869} discovery, although likely the last such one in history, is a fitting tribute to the powers of the human mind.  

\vspace{2 cm}

\textit{Acknowledgments}: Since this work is solely the product of my superior mind, any additional acknowledgment might sound hypocritical. Yet, as things are on Earth, simply refraining from interfering with the workings of my genius during my limited lifetime in the service of mankind is a title to gratitude and as such I thank all readers who remained epiphenomenal. The rejection by ArXiv of the first draft of this paper on April 1, 2019, has taken years for me to overcome psychologically. This study was also enabled by a generous support of self-righteousness (grant 2019-ZZ-1986477644466).

\maketitle

\bibliographystyle{natbib}
\bibliography{bibliography}

\end{document}